\documentclass[runningheads]{llncs}
\usepackage{graphicx}
\usepackage{listings}
\usepackage{makecell}
\usepackage{wrapfig}
\usepackage{lscape}
\usepackage[format=plain, belowskip=-20pt,aboveskip=2pt,labelfont=bf,textfont=it]{caption}

\begin{document}
\title{Exploring the acceleration of the Met Office NERC Cloud model using FPGAs}
\titlerunning{Exploring the acceleration of MONC using FPGAs}
%
\author{Nick Brown}
\authorrunning{N. Brown}
%
\institute{EPCC, The University of Edinburgh, Bayes Centre, Edinburgh, UK \\
\email{n.brown@epcc.ed.ac.uk}}
\maketitle              
\begin{abstract}
The use of Field Programmable Gate Arrays (FPGAs) to accelerate computational kernels has the potential to be of great benefit to scientific codes and the HPC community in general. With the recent developments in FPGA programming technology, the ability to port kernels is becoming far more accessible. However, to gain reasonable performance from this technology it is not enough to simple transfer a code onto the FPGA, instead the algorithm must be rethought and recast in a data-flow style to suit the target architecture. In this paper we describe the porting, via HLS, of one of the most computationally intensive kernels of the Met Office NERC Cloud model (MONC), an atmospheric model used by climate and weather researchers, onto an FPGA. We describe in detail the steps taken to adapt the algorithm to make it suitable for the architecture and the impact this has on kernel performance. Using a PCIe mounted FPGA with on-board DRAM, we consider the integration on this kernel within a larger infrastructure and explore the performance characteristics of our approach in contrast to Intel CPUs that are popular in modern HPC machines, over problem sizes involving very large grids. The result of this work is an experience report detailing the challenges faced and lessons learnt in porting this complex computational kernel to FPGAs, as well as exploring the role that FPGAs can play and their fundamental limits in accelerating traditional HPC workloads.

\end{abstract}
\section{Introduction}
The Met Office NERC Cloud model (MONC) \cite{easc} is an open source high resolution modelling framework that employs Large Eddy Simulation (LES) to study the physics of turbulent flows and further develop and test physical parametrisations and assumptions used in numerical weather and climate prediction. As a major atmospheric model used by UK weather and climate communities, MONC replaces an existing model called the Large Eddy Model (LEM) \cite{lem} which was an instrumental tool, used by scientists, since the 1980s for activities such as development and testing of the Met Office Unified Model (UM) boundary layer scheme \cite{lock1998}, convection scheme \cite{petch2001} and cloud microphysics \cite{hill2014}. In order to further the state of the art, scientists wish to model at a greater resolution and/or near real time which requires large amounts of computational resources. The use of modern HPC machines is crucial, however the problems are so challenging that any opportunity to accelerate the model is important. Whilst MONC has traditionally been run across thousands of Intel CPU cores in modern supercomputers \cite{easc}, a key question is what sort of architecture is optimal going forwards, and what changes are required to the code?

The idea of converting an algorithm into a form that can program a chip directly, and then executing this at the electronics level, has the potential for significant performance and energy efficiency advantages in contrast to execution on general purpose CPUs. However, the production of Application Specific Integrated Circuits (ASICs) is hugely expensive, and so a middle ground of Field Programmable Gate Arrays (FPGAs) tends to be a good choice. This technology provides a large number of configurable logic blocks sitting in a sea of configurable interconnect, and the tooling developed by vendors supports programmers converting their algorithms down to a level which can configure these fundamental components. With the addition of other facets on the chip, such as fast block RAM (BRAM), Digital Signal Processing (DSP) slices, and high bandwidth connections off-chip, FPGAs are hugely versatile. It's a very exciting time for this technology because, whilst they have a long heritage in embedded systems and signal processing, more recently there has been significant interest in using them more widely, such as the deployment of FPGAs in Amazon's F1 cloud computing environment

However, the use of FPGAs in scientific computing has, until now, been more limited. There are a number of reasons for this, not least the significant difficulty in programming them. But recent advances in high level programming tools means that this technology is now more accessible for HPC application developers. However it isn't enough to simply copy some code over to the FPGA tooling and synthesise it, a programmer must also change the entire way in which they approach their algorithms, moving to a data-flow style \cite{dataflow}, in order to achieve anywhere near good performance.

In this paper we describe the work done porting the computationally intensive advection kernel of the MONC atmospheric model to FPGAs. We compare this against the performance one can expect from more traditional Intel based CPU systems, and explore the many options and pitfalls one must traverse in order to obtain good performance of codes on FPGAs. In short, the contributions of this paper are

\begin{itemize}
    \item Exploration of the steps required to port a computationally intensive kernel onto FPGAs using HLS. We will show that it is not enough to simply copy the code over, but instead the whole approach needs to be rethought and recast.
    \item An experience report of using FPGAs to solve a computationally intensive kernel on large grids. We run experiments up to grid cells of 257 million grid points, each point requiring over fifty double precision operations.
    \item A detailed performance comparison, for this application, of the performance characteristics of our FPGA accelerated kernel in comparison to running on Intel CPUs commonly found in HPC machines. We are looking to answer the question, is it worth fitting Intel based systems with FPGAs?
\end{itemize}

This paper is structured as follows, in Section \ref{sec:bg} we describe the general background, introducing the MONC model in more detail, the FPGA hardware we are using in this work and describe the approach we have adopted in terms of programming the FPGA. In Section \ref{sec:HLS} we describe the development of our FPGA kernel, in HLS, and explore the different steps that were required to obtain reasonable performance from this code. In Section \ref{sec:blockdesign} we explore the block design adopted to integrate our kernel with the wider infrastructure supporting it. A performance comparison of our FPGA solution against Intel CPU products commonly found in HPC machines is explored in Section \ref{sec:performance}, before we draw conclusions and discuss further work in Section \ref{sec:conclusions}.

\section{Background}
\label{sec:bg}
\subsection{Met Office NERC atmospheric model}
\begin{wrapfigure}{r}{0.5\textwidth} 
\centering\raisebox{0pt}[\dimexpr\height-2.6\baselineskip\relax]{
\includegraphics[scale=0.6]{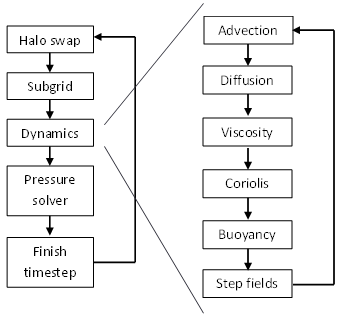}}
\caption{High level structure of a single MONC timestep}
\label{fig:tsstructure}
\end{wrapfigure}
The Met Office NERC Cloud model (MONC) has been developed in Fortran 2003 and, like many LES models, proceeds in timesteps, gradually increasing the simulation time on each iteration until reaching a predefined termination time. The model works on prognostic fields, \emph{u}, \emph{v} and \emph{w} for wind in the X, Y and Z dimensions, and a number of other fields which we do not considered in this paper. Figure \ref{fig:tsstructure} illustrates the high level structure of a timestep, where each piece of functionality executes sequentially, one after another. Initially, all prognostic fields are halo swapped between neighbouring processes and then the sub-grid functionality determines model parameterisations. Next, the dynamics group, often referred to as the dynamical core, performs Computational Fluid Dynamics (CFD) in order to solve modified Navier-Stokes equations, which is followed by the pressure solver, solving the Poisson equation. The timestep then concludes with some miscellaneous functionality such as checking for model termination. Over 70\% of the runtime is spent in the dynamical core and, in particular, the advection scheme. Advection calculates movement of values through the atmosphere due to wind, and at around 50\% of the overall runtime, is the single longest running piece of functionality. A number of different advection schemes are provided, and these require all the model's prognostic fields in order to complete their computation. 

In this work we concentrate on the Piacsek and Williams \cite{pwadvection} advection scheme which accounts for around 40\% of the model runtime. Listing \ref{lst:pw_orig} illustrates the MONC Fortran PW advection for advecting the \emph{u} variable with this scheme. Although this kernel also advects the \emph{v} and \emph{w} fields inside the same loop, details for those fields are omitted from Listing \ref{lst:pw_orig} for brevity as the calculations involved are very similar to those for advecting \emph{u}. It can be seen that the kernel is composed of three loops, each representing a dimension of our 3D space, and the inner loop, \emph{k}, in dimension Z loops up a single column. Starting at the second element of the column, this calculates contributions to the \emph{source term} of the flow field, \emph{su}, based upon values held in \emph{u}, \emph{v} and \emph{w}. Later on in the timestep, after the dynamical core has run, the source terms are then integrated into the flow fields. This advection kernel is a stencil based code, of depth one, accessing values of the three flow fields across all the three dimensions. All calculations are performed in double precision. 

\begin{lstlisting}[frame=lines,caption={Illustration of the PW advection scheme for the u field only},label={lst:pw_orig}, numbers=left]
do i=1, x_size
  do j=1, y_size
    do k=2, z_size
      su(k, j, i) = tcx * (u(k,j,i-1) * (u(k,j,i) + u(k,j,i-1)) - u(k,j,i+1) * (u(k,j,i) + u(k,j,i+1)))
      
      su(k, j, i) = su(k, j, i) + tcy * (u(k,j-1,i) * (v(k,j-1,i) + v(k,j-1,i+1)) - u(k,j+1,i) * (v(k,j,i) + v(k,j,i+1)))
      
      if (k .lt. z_size) then
        su(k, j, i) = su(k, j, i) + tzc1(k) * u(k-1,j,i) * (w(k-1,j,i) + w(k-1,j,i+1)) - tzc2(k) * u(k+1,j,i) * (w(k,j,i) + w(k,j,i+1))
      else 
        su(k, j, i) = su(k, j, i) + tzc1(k) * u(k-1,j,i) * (w(k-1,j,i) + w(k-1,j,i+1))
      end if
    end do
  end do
end do
\end{lstlisting}

\subsection{Hardware setup}
For the work described in this paper we are using an ADM8K5 PCI Express card, manufactured by Alpha Data, which mounts a Xilinx Kintex Ultrascale KU115-2 FPGA. This FPGA contains 663,360 LUTs, 5520 DSP48E slices and 4320 BRAM 18K blocks. The card also contains two banks of 8GB DDR4-2400 SDRAM, external to the FPGA, and a number of other interfaces which are not relevant to this work. Because the FPGA used for this work is part of the Xilinx product family, it is their general ecosystem, including tooling, that we use in this work. However, we believe that the lessons learnt apply more generally to product families of FPGAs from other vendors too.

This PCIe card is plugged into an Intel Xeon system, which contains two Sandybridge CPUs, each with four physical cores running at 2.40GHz, and 32GB RAM (16GB per NUMA region). Our approach is to run MONC on the CPU and study the benefit of offloading the PW advection scheme onto the PCIe mounted FPGA. Not only does this involve performing the double precision calculations for all three fields illustrated in Listing \ref{lst:pw_orig}, but also transferring the necessary flow field data onto, and resulting source terms back from, the card. Whilst some FPGAs such as the Zynq use a more embedded style, where typically ARM cores are combined with FPGA fabric on the same chip, we believe this PCIe setup is more interesting in the field of HPC. There are a number of reasons for this, firstly because a powerful Xeon CPU can be used on the host side, secondly because a large amount of memory can be placed close to the FPGA on the PCIe card to handle the processing of large problems, and thirdly because this is a very common accelerator architecture already adopted in HPC GPU systems.

\subsection{FPGA programming techniques and our approach}
The traditional approach to programming FPGAs has been to develop codes in a Hardware Description Language (HDL) such as VHDL or Verilog. However, this is a very time consuming process \cite{rtl-hard} which requires significant expertise and experience. As such, higher level programming tools have been developed to assist in programming, and High Level Synthesis (HLS) is amongst the most prevalent of these. A kernel, written in C, C++ or System C, is automatically translated, via HLS, into the underlying HDL. Driven by pragma style hints provided by the programmer, this substantially speeds up development time and allows for application developers to take advantage of the knowledge and experience of the FPGA vendor. An example of this is in floating point operations, where HLS will automatically include pre-built floating point cores to perform operations, instead of the HDL developer having to develop their own solution. 

HLS can be used as a building block of further programming abstractions, and recently the use of OpenCL for programming FPGAs has become popular \cite{opencl}. Decorating their code via OpenCL abstractions, a tool-chain such as Xilinx's SDAccel converts this into a form understandable by HLS, then uses HLS to generate the appropriate HDL and integrates this into a wider design based upon a board specific support package. An alternative approach which, in contrast to OpenCL, requires a bit more work on behalf of the programmer, is the use of the high-level productivity design methodology \cite{highproduct}. In this technique, the FPGA is configured using a block design approach, where existing IP blocks are imported and connected together by the programmer. This emphasises the reuse of existing IP and the idea of a \emph{shell}, providing general foundational functionality that the programmer's kernel, via an IP block generated by HLS, can be dropped into and easily integrated. By separating the shell from the kernel, the general shell infrastructure can be reused for many different applications, and updating the functionality of the kernel, which is quite common during development, often just requires re-importing the IP block into the board design. This approach also eases testing, as the kernel and shell can be validated separately. 

In this work we followed the high-level productivity design methodology, where one explicitly writes a C kernel for HLS, generates the HDL and export this as an IP block. This block is then integrated with a shell, providing the general infrastructure. There were two reasons for adopting this approach, firstly because it gave us more control over the configuration of our design, and we think that some of the lessons learnt for HPC codes could then potentially feed into higher level abstractions, such as those provided by OpenCL. Secondly, we chose this approach because SDAccel, the implementation of OpenCL for Xilinx FPGAs, is an extra commercial product that requires further licencing.

There have been a number of previous activities investigating the role that FPGAs can play in accelerating HPC codes. One such example is \cite{lfricfpga}, where the authors investigated using the high-level productivity design methodology to accelerate solving of the Helmholtz equation. They offloaded the matrix-vector updates requires as part of this solver onto a Zynq Ultrascale, however the performance they observed was around half of that when the code was run on a twelve core Broadwell CPU. Crucially, in our work, we are focused on accelerating a much more complicated kernel. In \cite{lfricfpga} the author's matrix-vector kernel involved looping over two double precision floating point operations, whereas in comparison the kernel we are offloading to the FPGA comprises of fifty three double precision floating point operations, twenty one double precision additions or subtractions, and thirty two double precision multiplications. We are also running on much larger grid sizes, and whereas in \cite{lfricfpga} the authors were limited to a maximum data size of 17MB due to keeping within the BRAM on the Zynq, in the work detailed in this paper we consider grid sizes resulting in 6.44GB of prognostic field data (and a further 6.44GB for the field source terms), necessitating the use of external SDRAM on the PCIe card.

\section{Developing the PW advection HLS kernel}
\label{sec:HLS}

\addtolength{\tabcolsep}{3pt}
\vspace*{-0.5cm}
\begin{figure}
\begin{tabular}{ | c c c c c | }
\hline
\textbf{Kernel description} & \makecell{\textbf{Runtime} \\ \textbf{(ms)}} & \makecell{\textbf{LUT} \\ \textbf{usage}} & \makecell{\textbf{DSP48E} \\ \textbf{usage}} & \makecell{\textbf{BRAM-18K} \\ \textbf{usage}} \\ \hline
Reference on CPU & 676.4 & NA & NA & NA\\ \hline
Initial port & 51498 & 9743 & 85 & 0 \\ \hline
Pipeline directive on inner loop & 14130 & 11356 & 58 & 64 \\ \hline
Local BRAM for column data & 3213.2 & 27598 & 267 & 130 \\ \hline
Local BRAM batches columns in Y & 1513.2 & 37474 & 393 & 453 \\ \hline
Extract all variables & 1301.6 & 38393 & 469 & 312 \\ \hline
Burst mode on port & 1097.2 & 40913 & 469 & 324 \\ \hline
Re-order X and Y loops  & 621.3 & 41151 & 469 & 324 \\ \hline
Replace memcpy with explicit loops  & 568.1 & 40638 & 466 & 324 \\ \hline
\makecell{Tune double precision cores \\ and clock to 310Mhz} & 514.9 & 27601 & 406 & 324 \\
\hline
\end{tabular}
\caption{Runtime of the PW advection kernel alone for different steps taken when porting it to HLS for a problem size of x=512,y=512,z=64 (16.7 million grid cells)}
\label{fig-kernel-optimisation}
\end{figure}
\addtolength{\tabcolsep}{-3pt}

Figure \ref{fig-kernel-optimisation} illustrates the performance of our HLS PW advection kernel over numerous steps that we applied one after another, for an experiment of x=512, y=512, z=64 (16.7 million grid cells). We use this table to drive our discussions in this section about the different steps required to optimise an HLS kernel and, for reference, the top entry of Figure \ref{fig-kernel-optimisation}, \emph{Reference on CPU}, illustrates the kernel's runtime over a single Sandybridge CPU core on the host. Our focus in this section is the HLS kernel alone, and as such we ignore the transferring of data which must occur to the PCIe before the kernel runs and the copying back of data once the kernel has completed. This transferring is considered in more detail in the next section. 

As a first step we ported the Fortran PW advection kernel for advecting the \emph{u}, \emph{v} and \emph{w} flow fields, (see Listing \ref{lst:pw_orig}) into C and the initial port of this onto the FPGA involved little more than just copying the C into the HLS tool and applying the correct directives to determine the type of external port interfaces. This is illustrated in Listing \ref{lst:initialhls_interface}, for the double precision array \emph{u}, representing wind in the X dimension, and source term \emph{su} which containing advected values calculated by this kernel for the X dimension. There are also arrays for \emph{v}, \emph{sv}, \emph{w} and \emph{sw} which are omitted from Listing \ref{lst:initialhls_interface} due to brevity. In this section we refer to these six arrays as the \emph{kernel's external data arrays}, and these use the AXI4 protocol. The \emph{offset=slave} specifier bundles them into a single port and instructs HLS that these point to different addresses in the control bus port. Other scalar variables such as the size of the data in each dimension, \emph{size\_x} and \emph{size\_y} in Listing \ref{lst:initialhls_interface}, are also provided via the AXI4-Lite control bus port. This control bus port is exposed to the host as a memory block, which the host can then write to, and HLS provides the explicit location in this block of the different variables. This initial version was trivial, both from a code and FPGA utilisation perspective, requiring less than 10,000 LUTs, a handful of DSP48E slices and no BRAM, however at 51498 milliseconds (51 seconds) it was very slow. 

\begin{lstlisting}[frame=lines,caption={Skeleton of HLS main function, illustrating external data interfaces},label={lst:initialhls_interface}, numbers=left, language=c]
int pw_advection(double * u, double * su, ..., int size_x, int size_y, ...) {
    #pragma HLS INTERFACE m_axi port=u offset=slave 
    #pragma HLS INTERFACE m_axi port=su offset=slave

    #pragma HLS INTERFACE s_axilite port=size_x bundle=CTRL_BUS
    #pragma HLS INTERFACE s_axilite port=size_y bundle=CTRL_BUS
    #pragma HLS INTERFACE s_axilite port=return bundle=CTRL_BUS
    .....
}
\end{lstlisting}

We next added the HLS pipeline directive, \emph{\#pragma HLS PIPELINE II=1}, to our inner loop, working up a single column. This instructs HLS to pipeline the processing of the inner loop and \emph{II} is the initiation interval latency, which instructs HLS how often to add a new element of data to the pipeline, in this case every cycle if possible. HLS takes the inner loop, containing our 53 double precision operations, and breaks this up into individual pipeline stages which can run concurrently. These are then fed with data from the outer loops and, once the pipeline is filled, each stage is running, at the same time, on a different element of data before passing the result to the next stage. From Figure \ref{fig-kernel-optimisation} it can be seen that this significantly decreased the runtime of the kernel, by around five times, but this was still around twenty time slower than running on the CPU.

The bottleneck at this stage was that the data port, used to implement the kernel's external data arrays, that we read from extensively in the calculation of a single column, does not support more than one access per clock cycle. Hence we were instructing HLS to pipeline the inner loop, so that functionality within it runs concurrently on different elements of data. But crucially HLS realised that there would be numerous conflicts on the data port if it fully pipelined the calculations, and as such very severely limited the design of the pipeline. To address this, we created a number of local arrays to hold all the data required for working with a single column, and this is illustrated in Listing \ref{lst:working_arrays}. We actually created twenty two arrays, for the six kernel external data arrays, three input flow field and three output source term arrays, which included columns in other X and Y dimension indexes. Threes of these local arrays, \emph{u\_vals}, \emph{u\_xp1\_vals} (holding a column of data of the X+1 column) and \emph{u\_vals2} are illustrated in Listing \ref{lst:working_arrays} and these are set to a static size (\emph{MAX\_VERTICAL\_SIZE}) as array sizes can not be dynamically sized in HLS. The arrays are filled with the data required for a column via the memory copies at lines 5 to 7 (note \emph{u(i,j,0)} is a preprocessor directive that expands out to index the appropriate 3D location in the array), before executing the calculations needed on that column. The idea was that all accesses on the port are before the calculations start and therefore there are no conflicts during the pipelined inner loop. We use on-chip BRAM to store these array, and HLS can make these either single ported or dual ported, meaning that it can either be accessed once or twice independently in a clock cycle. The challenge here is that, for a specific column, there are a large number of accesses to each array within the inner loop, for instance there are nine accesses to the \emph{u\_vals} array which holds the current column's data for the \emph{u} external data array.

\begin{lstlisting}[frame=lines,caption={Using local BRAM to store data for a single column data},label={lst:working_arrays}, numbers=left, belowcaptionskip=-3pt, language=c]
double u_vals[MAX_VERTICAL_SIZE], u_xp1_vals[MAX_VERTICAL_SIZE], u_vals2[MAX_VERTICAL_SIZE], ....;

for (unsigned int i=start_x;i<end_x;i++) {
    for (unsigned int j=start_y;j<end_y;j++) {
        memcpy(u_vals, &u(i,j,0), sizeof(double) * size_z);
        memcpy(u_xp1_vals, &u(i+1,j,0), sizeof(double) * size_z);
        memcpy(u_vals2, &u(i,j,0), sizeof(double) * size_z);
        ....
        for (unsigned int k=1;k<size_z;k++) {
        #pragma HLS PIPELINE II=1
            .....
        }
    }
}
\end{lstlisting}

To address this we duplicated these same arrays, for instance \emph{u\_vals2} which holds the same data as \emph{u\_vals}. Whilst another way around this in HLS is to use partitioning, effectively splitting the array up across multiple BRAM controllers, due to the dynamic size of the inner loop, we would have been forced to partition the arrays into single elements, and this resulted in worse utilisation and performance. In comparison, duplicating the BRAM array worked well.

It can be seen from Figure \ref{fig-kernel-optimisation} that this local copying of data decreased the runtime by over four times. However, the major disadvantage of the approach in Listing \ref{lst:working_arrays} is that the outer loops of \emph{j} in the Y dimension and \emph{i} in the X dimension are no longer continually feeding data into the pipelined inner loop. Instead, the inner loop runs in a pipelined fashion just for a single column, in this case of maximum 64 data elements, and then must drain and stop, before memory copies into local arrays are performed for the next column. Bearing in mind the pipeline of this inner loop is, as reported by HLS, 71 cycles deep, with an initiation interval, the best HLS can provide, of 2 cycles and assuming a column size of 64 elements, for each column the pipeline will run for 199 cycles but for only 57 of these cycles (28\%) is the pipeline full utilised, the rest of the time it is either filling or draining.

To address this we extended our local BRAM arrays to hold data for multiple columns in the Y dimension, extending each array from \emph{MAX\_VERTICAL\_SIZE} to \emph{MAX\_VERTICAL\_SIZE * Y\_BATCH\_SIZE}. In this situation the middle loop, \emph{j}, working in the Y dimension runs in batches, of size \emph{Y\_BATCH\_SIZE}. For each batch it will copy the data for \emph{Y\_BATCH\_SIZE} columns, and then process each of these columns. The major benefit of this approach is that our pipeline, working up the column in the inner loop, is now fed by \emph{Y\_BATCH\_SIZE} columns rather than one single column. Additionally, at this point, HLS reported that it had been able to reduce the initiation interval down from two to one, effectively doubling the performance of the inner loop pipeline. Assuming a \emph{Y\_BATCH\_SIZE} of 64 and that the column size is still 64, the pipeline now runs for 4167 cycles, 97\% of which the pipeline is fully filled. This represents a significant increase in utilisation, and ultimately performance, because the pipeline is able to process, and hence generate a result, every clock cycle for 97\% of the time it is running. As per Figure \ref{fig-kernel-optimisation}, this over halved the kernel execution time at the cost of increasing the BRAM usage by over three times.

At this point the individual lines of code for our inner loop kernel, containing the fifty three double precision floating point operations, were still laid out similarly to Listing \ref{lst:pw_orig}, where the calculations for a specific value of the source term were in one line. We had trusted HLS to extract out the individual variable accesses, and structure these appropriately, but we found that actually HLS does a fairly poor job of identifying which variables are shared and hence can be reused between calculations. As such, we significantly restructured the code, splitting up calculations into their individual components of reading data into temporary variables and then using that single variable whenever the value is required in the inner loop. This is the \emph{Extract all variables} entry of Figure \ref{fig-kernel-optimisation} and had two impacts. Firstly, it reduced the pipeline depth from 71 cycles deep to 65, and hence provided a modest increase in performance, but also it reduced the number of reads on our local arrays and so we were able to remove a number of duplicate local arrays which reduced the overall BRAM usage by around 30\%.

When issuing memory copies, for instance in lines 5 to 7 of Listing \ref{lst:working_arrays}, the port must be read which accesses data from external SDRAM, and the same is true in the other direction when writing data. In the default mode, ports will tend to issue an access for every individual element, but instead it is possible to decorate the kernel's external data array variable definitions (e.g. \emph{u} and \emph{su}) with pragmas to instruct HLS to issue bursts of data, retrieving \emph{n} elements in one go. This is important in our situation because we are never just reading one element at a time, but instead the data for \emph{Y\_BATCH\_SIZE} columns. The \emph{HLS INTERFACE} pragma, as illustrated in Listing \ref{lst:initialhls_interface}, was modified for our kernel's external data arrays (e.g. \emph{u} and \emph{su}) with the addition of \emph{num\_read\_outstanding=8 num\_write\_outstanding=8 max\_read\_burst\_length=256 max\_write\_burst\_length=256}. This directs HLS to read and write in bursts of size \emph{256}, the maximum size, and supports holding up to eight of these bursts at any one time. The \emph{latency} modifier advises HLS to issue the access before it is needed, in this example around 60 cycles beforehand. HLS uses BRAM to store these bursts and, as can be seen in Figure \ref{fig-kernel-optimisation}, resulted in a modest increase in BRAM usage but also a reasonable decrease in execution time.

At this point we are working in batches of columns and as our middle, \emph{i}, loop running over the Y dimension, reaches the limit of one batch it stops and retrieves data from memory for the next batch. Crucially, this happens for every iteration in the outer loop, \emph{i}, over the X dimension and as the code progresses from one level in X to the next, then all batches in Y are run again. The problem with this is that there are fifteen memory copies required for every batch and this involves significant amounts of time accessing the DRAM. It is possible to address this by moving the outer loop, \emph{i}, over the X dimension, inside the \emph{j} middle loop which is running over a single batch of columns. This means that memory accesses themselves in the X dimension are effectively pipelined too, and is illustrated by Listing \ref{lst:pipelineing_mem}. For brevity we just show a subset of the variables and local arrays, but it is enough to demonstrate the approach. It can be seen that the loop ordering has changed, such that the outer loop is now looping \emph{m} times, once per batch of \emph{MAX\_Y\_SIZE} columns at line 1. The start of a batch requires data to be copied into the \emph{up1\_vals} variable, representing the column plus one in the X dimension. Then, as the loop progresses through levels in the X dimension, for each next level, \emph{u\_vals} is populated with data from \emph{up1\_vals} and only the plus one level in the X dimension, i.e. \emph{up1\_vals} needs to go to SDRAM memory to retrieve the \emph{i+1} column in X. All other copies, for instance \emph{u\_vals} at line 5, are accessing chip local BRAM which is much faster than going off chip to the SDRAM. This significantly reduces the number of off chip data accesses to DRAM that need to be performed and almost halves the runtime of the kernel.

\begin{lstlisting}[frame=lines,caption={Reordering the X and Y loops to pipeline memory access in the X dimension},label={lst:pipelineing_mem}, numbers=left, belowcaptionskip=-3pt, language=c]
for (unsigned int m=start_y;m<end_y;m+=MAX_Y_SIZE) {
    memcpy(up1_vals, &u(start_x,m,0), sizeof(double) * MAX_VERTICAL_SIZE*MAX_Y_SIZE);
    ....
    for (unsigned int i=start_x;i<end_x;i++) {
        memcpy(u_vals, up1_vals, sizeof(double) * MAX_VERTICAL_SIZE*MAX_Y_SIZE);
        memcpy(up1_vals, &u(i+1,m,0), sizeof(double) * MAX_VERTICAL_SIZE*MAX_Y_SIZE);
        ....
        for (unsigned int j=0;j<MAX_Y_SIZE;j++) {
            ....
        }
    }
}
\end{lstlisting}

Until this point we have relied on the use of the \emph{memcpy} function to copy data from one location to another. However bearing in mind there are multiple copies of local column data arrays due to BRAM port limits, e.g. \emph{u\_vals} and \emph{u\_vals2}, issuing a separate memcpy for each of these when we loop into the next X dimension is quite slow because HLS will not execute these memory copies concurrently. Instead, replacing the \emph{memcpy} calls with explicit loops, where each index location is read from the source array and then written to each of the target arrays was faster. In fact, more generally we found that replacing all the \emph{memcpy} calls with an explicitly pipelined loop that performed the copying in user code, was beneficial. This is represented as the \emph{Replace memcpy with explicit loops} entry of Figure \ref{fig-kernel-optimisation} and it can be seen that not only did we obtain a modest increase in performance, but it also decreased our LUT utilisation slightly.

The default clock on the ADM8K5 board is 250Mhz, and so a period of 4ns was used initially, with HLS estimating a clock period of 3.75ns due to limits in double precision multiplication. However, via configuring the HLS floating point kernels it was possible to tune them. Using \emph{\#pragma HLS RESOURCE variable=a core=DMul\_maxdsp latency=14}, HLS was instructed to use the \emph{DMul\_maxdsp} double precision floating point core (leveraging DSP slices as much as possible for the double precision multiplication) with a latency of 14 cycles for all multiplications involving the variable \emph{a}. This latency is the core's pipeline depth and, by increasing it, it is possible to reduce the minimum clock period. We applied this directive to all variables that are involved in double precision multiplication, and found that the best clock period we could get from the double precision multiplication core was 2.75ns. Whilst the latency value can go all the way up to twenty, above fourteen made no difference to the period. As such we were able to reduce our clock period to 3.2 (there is a 12.5\% clock uncertainty), meaning we could run our kernel at 310Mhz instead of 250Mhz. The pipeline depth has increased from 65 to 72, but due to the increase in clock frequency, the overall latency for data to progress through the pipeline has gone from 2.6e-7 seconds to 2.3e-7 seconds, so there is an increase in overall performance. 

From Figure \ref{fig-kernel-optimisation} it can be seen that the LUT and DSP48E utilisation dropped very significantly with this last configuration. This was because we also instructed HLS to use the full DSP core when it came to double precision addition and subtraction. It is the use of this core that reduced the LUT usage by around a quarter, and also, ironically, slightly reduced the number of DSP48E slices too. 

As a result of the steps applied in this section, we have reduced the runtime of our HLS kernel by over 100 times, from being 75 times slower than running over a single Sandybridge CPU core on the host, to being around a quarter faster, however as noted at the start of the section this is just the kernel execution time and ignores the DMA transfer time needed to get data on and off the board before and after the the kernel runs.

\section{Putting it all together, the block design}
\label{sec:blockdesign}

\begin{landscape}

\begin{figure}
\includegraphics[scale=0.48]{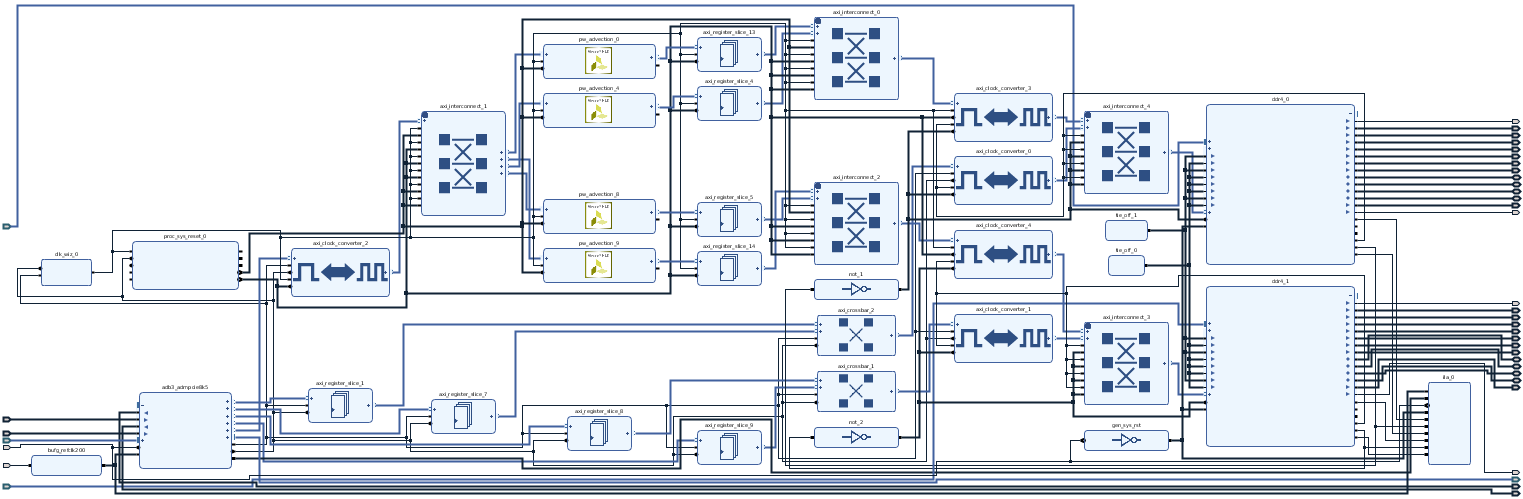}
\caption{The MONC PW advection board design, containing four of our PW advection HLS kernels and other general infrastructure to support this}
\label{fig:shell}
\end{figure}

\end{landscape}

Once developed, we then need to integrate our PW advection kernel with general infrastructure to connect our kernel to the PCIe interface and on card SDRAM. The general workflow is that field data is transferred from the host to the on card SDRAM via DMA and kernels are then run, reading data from the SDRAM and writing source term results to SDRAM, and once complete results are transferred back from the SDRAM to the host via DMA. Figure \ref{fig:shell} illustrates the block design of our system, and in this design we are using four PW advection kernels, towards the top centre of Figure \ref{fig:shell} with the HLS logos. The IP block on the bottom left is the PCIe interface, providing four independent DMA channels that can be used for communication and a direct slave interface that can also be used for communication. The two big blocks on the far right are memory controllers for the on card SDRAM, each controller responsible for one of the two banks of on card 8GB memory. We connect the first two PCIe interface DMA channels to the first memory controller, and the other two DMA channels to the second memory controller. In between the PCIe interface and their corresponding SDRAM memory controller, these connections pass through infrastructure which, for instance, converts the clock from the PCIe clock domain to the SDRAM memory controller clock domain. In this design, to avoid bottlenecks, the two banks of 8GB memory are entirely separate and it is not possible for an IP block connected to one bank to access memory of the other bank. 

\begin{figure}
\centering
\includegraphics[scale=0.38]{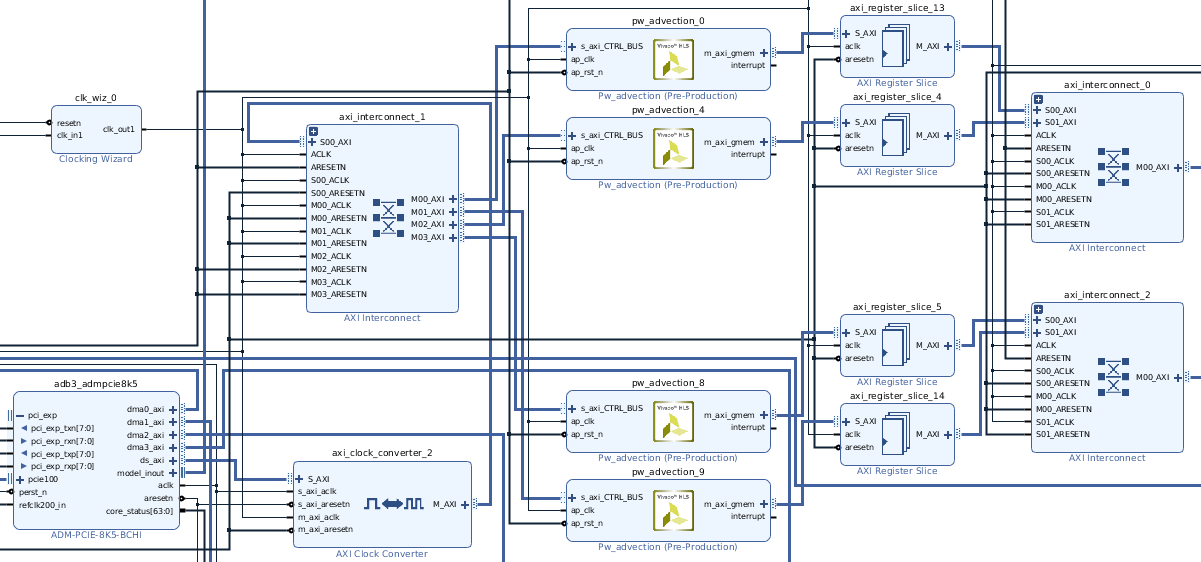}
\caption{Integration of the PW advection HLS IP blocks in our board design}
\label{fig:kernelzoom}
\end{figure}

Figure \ref{fig:kernelzoom} provides a more detailed view of the integration of our PW advection kernel IP blocks. For purposes of illustration, we have slightly moved the appropriate IP blocks around, when compared to Figure \ref{fig:shell}, so the topical ones are in the same image. On the bottom left of Figure \ref{fig:kernelzoom}, the \emph{ADM PCIe} block is the PCIe interface and the four DMA channel ports on the right of this IP block, along with the direct slave port can be clearly seen. The clocking wizard, top left, converts the 250Mhz reference clock up to 310Mhz for the PW advection IP blocks as described in Section \ref{sec:HLS}. The direct slave interface is connected to the kernel's control port, and by writing or reading the appropriate bit we can manage the kernels such as starting or tracking progress. This connection goes, via an AXI4 clock converter IP block, to the slave interface of the left most AXI interconnect. This interconnect splits the direct slave data according to its address, the appropriate data then routed to its corresponding PW advection kernel. These addresses are defined in the block design address editor.

The main data port, \emph{m\_axi\_gmem}, of the PW advection kernel is on the right of the PW advection IP block and it is through this port that the kernel's external data arrays such as \emph{u} and \emph{su} are routed. This AXI4 data port connects, via an AXI register slice, to an AXI interconnect on the right of Figure \ref{fig:kernelzoom}. Our PW advection kernels are split into two groups, one group connecting to the first SDRAM memory controller (one bank of 8GB RAM) and the second group connecting to the second DDR SDRAM memory controller (the second bank of 8GB RAM). This is the reason for the two AXI interconnects on the right, one for each group of kernels and we do it this way with the aim of reducing congestion. For purposes of illustration, in Figures \ref{fig:shell} and \ref{fig:kernelzoom} we limited ourselves to just four PW advection IP blocks. However our design is scalable and adding additional PW advection kernels just requires reconfiguration of the appropriate AXI interconnects to add more ports and assigning an address to the new IP block's control bus port.

When it came to the shell and overall kernel integration, we experimented with a number of different designs to understand which would provide the best performance. The main driver here is the speed to access the SDRAM and avoid memory accesses becoming a bottleneck. We found this design, where we keep the two banks of 8GB SDRAM entirely separate and connect each to two different DMA channels, to provide the best performance. A summary of these investigations is illustrated in Figure \ref{fig-shell-optimisation}, which describes the transfer time, via DMA, to copy 1.6GB of data from the host to the DRAM on the PCIe card. The first row of Figure \ref{fig-shell-optimisation} is the design that we have described in this section. The second row, \emph{one memory controller only}, is when all four DMA channels are used but we only copy the data into one of the memory controllers, and it can be seen that this increases the transfer time by around a fifth. The third entry is where there are two memory controllers, each serviced by two DMA channels, but these are connected together in one large memory space so any memory access can see all the 16GB SDRAM. This is slightly slower than our split approach, because all memory accesses go through a single AXI interconnect, but there isn't much in it. The last entry of the table, is where the two banks of memory is kept separate, but we only drive these via one single DMA channel. This adds around a third to the overall DMA transfer time because transfers on the same channel need to queue up and so being able to spread them out across multiple channels and transfer concurrently is optimal.

\begin{figure}
\vspace*{-0.5cm}
\begin{center}
\begin{tabular}{ | c c | }
\hline
\textbf{Description} & \makecell{\textbf{DMA transfer time} \\ \textbf{(milliseconds)}} \\ \hline
Design described here & 232 \\
One memory controller only & 280 \\
Two memory controllers connected & 239 \\
One DMA channel per memory controller & 342 \\
\hline
\end{tabular}
\end{center}
\caption{DMA transfer time for different configurations with a data size of 1.6GB}
\label{fig-shell-optimisation}
\end{figure}

\section{Performance comparison}
\label{sec:performance}
We built the block design described in Section \ref{sec:blockdesign} with twelve PW advection kernels as described in Section \ref{sec:HLS}. In order to fit within the limits of the Kintex FPGA we instructed HLS to use the medium DSP core for double precision multiplication. This resulted in an overall design utilisation of 78.5\% of the Kintex's LUTs, 84.2\% BRAM-18k blocks and 89\% of the chip's DSP48E slices. It took around fifteen hours of CPU time to build the entire design, most of which was spent in the place and route phases.

\begin{figure}
\centering
\includegraphics[scale=0.5]{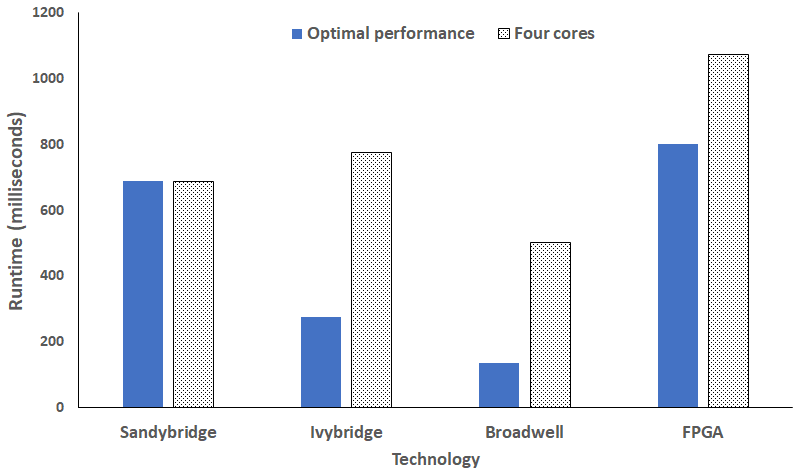}
\caption{Performance comparison of x=1012, y=1024, z=64 (67 million grid points) with a standard status cloud test-case}
\label{fig:performance-overall}
\end{figure}

Once built, we compared the performance of our PW advection FPGA design against a C version of the same PW advection algorithm, threaded via OpenMP across the cores of the CPU. For all runs the host code was compiled with GCC version 4.8 at optimisation level 3 and the results reported are averaged across fifty timesteps. Figure \ref{fig:performance-overall} illustrates a performance comparison of our FPGA kernel against the CPU only code running on Sandybridge, Ivybridge and Broadwell CPUs for a standard MONC stratus cloud test case of size x=1012, y=1024, z=64 (67 million grid cells). For each technology there are two runtime numbers, in milliseconds. The first, \emph{optimal performance}, illustrates the best performance we can get on the CPU by threading over all the physical cores (4 in the case of Sandybridge, 12 in the case of Ivybridge, 18 in the case of Broadwell, and 12 in the case of our FPGA design.) We also report a four core number, which includes only running over four physical cores, or PW advection kernels in the case of our FPGA design, as this is the limit of the Sandybridge CPU and it allows more direct comparison. 

It can be seen from Figure \ref{fig:performance-overall} that our optimal FPGA version performs slower in comparison to all CPU products tested and this trend continues when we consider four core performance. This might seem strange seeing that, in Section \ref{sec:HLS}, our HLS kernel is faster than running on a single core of Sandybridge and actually is comparable to running on a single core of Broadwell. However, crucially there we were just concerned with the kernel execution time and ignored the DMA transfer time and the results reported in Figure \ref{fig:performance-overall} contain both these aspects. 

\begin{figure}
\centering
\includegraphics[scale=0.6]{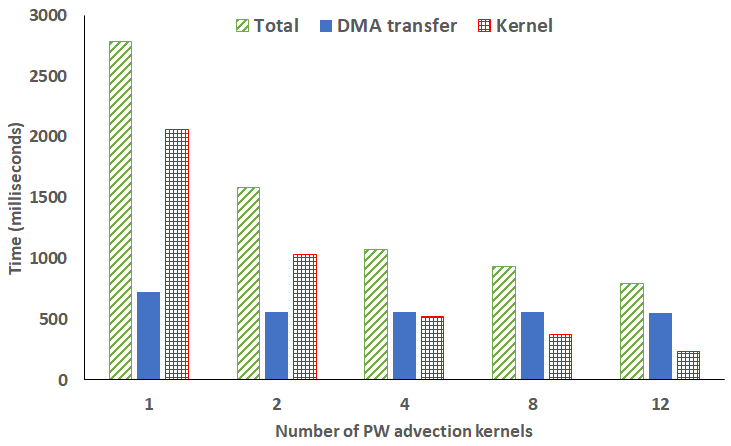}
\caption{Runtime of different numbers of PW advection kernel broken down into constituent parts when scaling number of kernels, x=1012, y=1024, z=64 (67 million grid cells) with a standard status cloud test-case}
\label{fig:performance-brokendown}
\end{figure}

To understand this further, Figure \ref{fig:performance-brokendown} illustrates the same experiment setup as we scale the number of PW advection kernels from one to twelve. For each configuration we report the total runtime and then break it down to the total time required for DMA transfer of data both to and from the card, and the kernel execution time. In all cases we distribute the cores as evenly as possible across groups, for instance two cores uses 1 core from each group, four cores uses two cores from each group. This experiment represents a grid size of 67 million points, and three fields, each point of which is double precision, hence a total of 3.32GB is being transferred. It can be seen that, for small numbers of advection kernels, the kernel runtime is by far the most significant. However this scales well and as we reach four kernels and beyond the DMA transfer time becomes dominant and 70\% of the total time with twelve kernels is in DMA transfer.

\begin{figure}
\centering
\includegraphics[scale=0.65]{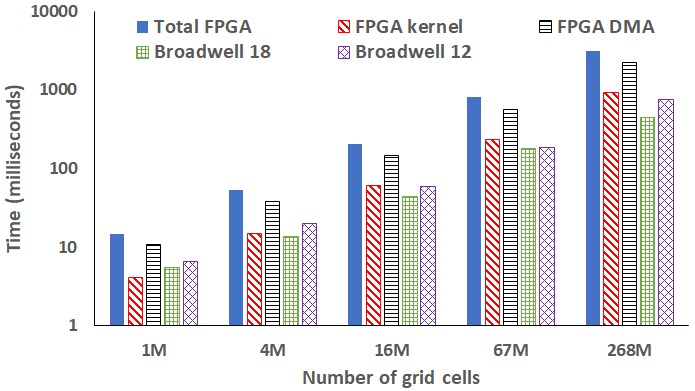}
\caption{Runtime of our FPGA PW advection code (12 kernels) vs Broadwell as we scale the grid size with a standard stratus cloud test-case}
\label{fig:scaling-grid}
\end{figure}

Figure \ref{fig:scaling-grid} illustrates how the time, in milliseconds, changes as we scale the number of grid cells, note that this is a log scale. We report three numbers for our FPGA approach (12 PW advection kernels), the total time, the execution time of the kernel and the total DMA transfer time. For comparison we also illustrate the runtime of the code on 18 physical cores of Broadwell and 12 physical cores of Broadwell (as we have 12 PW advection kernels). Whilst our PW advection FPGA version is slower than both Broadwell configurations, the FPGA HLS kernel itself is faster at 1 million grid cells, competitive with both at 4 million grid cells and competitive with the 12 Broadwell cores until 16 million grid cells. In terms of FLOPs, at 268 million grid cells our HLS kernel is providing 14.36 GFLOP/s (in comparison to 12 cores of Broadwell at 17.75 GFLOPs/), however when one includes the DMA transfer time this drops down to 4.2 GFLOP/s and so illustrates the very significant impact that DMA transfer time has on our results. The limit with some other investigations such as \cite{lfricfpga}, is that they focus on the embedded CPU-FPGA Zynq chip, and limit their system size very severely to the BRAM on that chip. As such they don't encounter this transfer time overhead, but this is crucially important to bear in mind for processing realistic problems that are of interest to scientists.

\section{Conclusions and further work}
\label{sec:conclusions}
In this paper we have described our approach in porting the PW advection kernel of the MONC atmospheric model onto FPGAs. Using HLS, we explored in detail the different steps required in porting the kernel and how best to structure the board design. We have shown that it is crucial that HPC application developers rethink and recast their code to suit the data-flow pattern, and have demonstrated a 100 times performance difference between code that does not do this and the same functionality, albeit where the code looks very different, tuned for the FPGA. This re-enforces the point that, whilst it is fairly simple for an HPC applications developer to import some C code into HLS and synthesis this, significant work and experimentation is required to get reasonable performance.

When considering only the kernel execution time, our HLS kernel outperformed a single core of Sandbridge and performs comparable with a single Broadwell core. But when including the DMA transfer time we found that this is a very severe limitation of our FPGA design in contrast to the performance one can gain from the same advection kernel threaded over the cores of Intel CPUs commonly found in HPC machines.

When it comes to further work, it is this DMA transfer time that needs to be tackled. At the largest problem size of 268 million grid cells explored in this paper, a total of 12.88GB needs to be transferred which takes 2.2 seconds and represents a transfer rate of 5.85 GB/s, which is reasonable based on the specifications of this PCIe card. One idea is to use single rather than double precision, which would effectively halve the amount of data that needs to be transferred, although the DMA transfer time at 134 million grid points is still substantially slower than Broadwell execution time at 268 million grid points. Another idea is to chunk up the DMA transfer, starting the appropriate PW advection kernel as soon as the applicable chunk has arrived rather than all the data, this could be driven by a host thread or extra logic in the block design.

There is also further exploration that can be done based on our HLS kernel and one aspect would be to look at how best to further increase the clock speed. The 2.8ns period of the double precision multiply currently limits us to 310Mhz, but replacing double precision with single or fixed point could enable higher clock frequencies and single precision would also halve the amount of data transferred from SDRAM to the kernels. This last point is important, because we believe that SDRAM access time is now the major source of overhead in our HLS kernel.

Therefore, we conclude that, whilst FPGAs are an interesting and generally speaking viable technology for accelerating HPC kernels, there is still work to be done to obtain performance competitive to modern day Intel CPUs. When running large system sizes on PCIe mounted FPGAs, the cost of transferring data to and from the card can be very significant and severely limits performance. This is important to bear in mind and, going forwards, if we can address these limits then we feel that FPGAs will become a much more competitive approach.

\section{Acknowledgements}
The authors would like to thank Alpha Data for the donation of the ADM8K5 PCIe card used throughout the experiments of work. This work was funded under the EU FET EXCELLERAT CoE, grant agreement number 823691.

%
%
%
%

\end{document}